\documentclass[a4paper,12pt]{article}
\usepackage{graphicx}
\usepackage{amssymb}
\usepackage{amsmath}
\usepackage{color}
\usepackage{hyperref}

\title{Simulation studies of hadron energy resolution as a function of iron plate thickness at INO-ICAL}

\author{
{\small S.M. Lakshmi,}$^{1}$ 
{\small A. Ghosh,}$^{2}$
{\small M. M. Devi,}$^{3}$
{\small D. Kaur,}$^{4}$ 
\and
{\small S. Choubey,}$^{2}$
{\small A. Dighe,}$^{3}$ 
{\small D. Indumathi,}$^{1}$ 
{\small M.V.N. Murthy}$^{1}$
and
{\small Md. Naimuddin}$^{4}$ \\
\vspace{0.03cm}
 \and 
$^{1}${\it \small The Institute of Mathematical Sciences, Chennai 600 113, India}
\and
$^{2}${\it \small Harishchandra Research Institute, Allahabad 211 002, India}
 \and 
$^{3}${\it \small Tata Institute of Fundamental Research, Mumbai 400 005, India}
 \and 
$^{4}${\it \small Department of Physics, Delhi University, New Delhi 110 007, India}
}

\begin{document}
\maketitle

\begin{abstract}
We report on a detailed simulation study of the hadron energy resolution as a
function of the thickness of the absorber plates for the proposed Iron
Calorimeter (ICAL) detector at the India-based Neutrino Observatory
(INO). We compare the hadron resolutions obtained with absorber
thicknesses in the range 1.5--8 cm for neutrino interactions in
the energy range 2--15 GeV, which is relevant to hadron production
in atmospheric neutrino interactions. We find that at lower energies,
the thickness dependence of energy resolution is steeper than at higher
energies, however there is a thickness-independent contribution
that dominates at the lower thicknesses discussed in this work. As a
result, the gain in hadron energy resolution with decreasing plate
thickness is marginal. We present the results in the form of fits to a
function with energy-dependent exponent.

\end{abstract}

\section{Introduction}\label{section1}

The proposed India-based Neutrino Observatory (INO) is an underground
laboratory designed primarily for the study of neutrinos from various
sources. One of the first and largest detectors at INO will be a magnetized
iron calorimeter (ICAL) detector to primarily study atmospheric muon
neutrinos (and anti-neutrinos). The ICAL detector is a calorimeter which
comprises a stack of iron plates interleaved with Resistive Plate Chambers
(RPC)\cite{ref:ino,ref:rpc} as active detector elements.

Muons produced in the charged-current (CC) interactions of atmospheric
muon (anti-)neutrinos with the iron target can be easily detected
in ICAL. Their energy and momentum will be reconstructed using either
the track length in the detector and/or the curvature due to the
magnetic field. The magnetic field allows the charge identification as
well.

The main difficulty in reconstructing the energy and direction
of the incident neutrino arises from the uncertainty in energy and
direction of the associated hadrons, mainly pions produced in interactions
at neutrino energies of $\gtrsim$ 1~GeV. Only hit
multiplicity in the RPC layers and its distribution are available in the
study of hadron response of ICAL \cite{ref:moon}. Furthermore, the hits are 
restricted to a few layers only in contrast to the long track of the minimum 
ionising muons.

A potentially crucial factor in the determination of hadron energy and
direction resolution is the thickness of the absorber material, i.e.,
iron. This must be optimised taking into account the physics goals of the
experiment, apart from the detector size, geometry, stability and cost. In
this simulation study, the effect of the variation in plate thickness on
hadron energy resolution is analysed using fixed energy
single pions. Earlier, such studies have concentrated on very high
energy hadrons from tens of GeV to hundreds of GeV in hadron calorimeters
\cite{ref:green,ref:dangreen,ref:Fabjan,ref:Kleinknecht}. These studies
have indicated a square root dependence on the thickness $t$ of the
hadron energy resolution on the absorber thickness. But no corresponding
systematic analysis of the absorber thickness dependence at lower
energies, in the GeV region, is found in the literature, although the
values of the hadron energy resolution at some fixed thicknesses are
available. Naively, hadrons traversing the plates at an
angle $\theta$ will ``encounter", in principle, an effective plate thickness
$(t/\cos\theta)$ so that the thickness dependence can be explored through
this angle dependence. However, in the actual detector, the detector
geometry including support structures, and orientation as well as the
arrangement of the detector elements introduce additional nontrivial
dependence on thickness. This is what this work intends to understand.

The main focus of this report is to present the results of a simulation study of
the thickness dependence of hadron energy resolution in the energy
range 2--15 GeV. This energy domain is of primary importance to
neutrino oscillations studies with ICAL. The default design of ICAL at
present uses iron plates of thickness 5.6 cm. The energy resolution of
hadrons (both for fixed energy single pions as well as for multiple
hadrons from CC interactions of neutrinos with iron) propagating in
the default configuration of the ICAL detector has been discussed in
Ref.~\cite{ref:moon}. Here, we study the effect of varying the plate
thickness in the range of 1.5 to 8 cm.

The report is structured as follows. In Section 2, the detector
configuration and the methodology of the analysis are outlined. The
energy dependence of the hadron energy resolutions with different
plate-thicknesses is discussed in Section 3. The results for the thickness
dependence of the resolution parametrised in the form $p_0 t^{p_1} +
p_2$ are presented in Section 4. Energy resolutions in
different bins of $\cos\theta$, where $\theta$ is the incident hadron
direction, and their thickness dependences are discussed in Section 5.
Section 6 compares ICAL simulations with test beam data and simulations
from both MONOLITH and MINOS. We conclude with a brief discussion and
summary in Section 7.

\section{Detector configuration and methodology}

The default configuration of ICAL detector has three modules of 151 
layers of 5.6 cm thick magnetized iron plates interleaved with RPCs; 
each module has a dimension of 16 m $\times$ 16 m $\times$ 14.45 m. 
The RPCs are placed in the 4 cm gap between two iron plates. Copper 
pickup strips of width 1.96 cm above and below each RPC, aligned
transverse to each other, determine the $x$ and $y$ coordinates of the
hit. The layer number gives the information about the $z$-coordinate. The
$x$-, $y$- and $z$- axes are defined with respect to an origin located at
the center of the detector, with the $x$ axis along the largest dimension
of the detector. For more details of the geometry and analysis, see
Ref.~\cite{ref:moon}.

A GEANT4-based \cite{ref:geant4} simulation framework has been used for
the current analysis. The strip width and spacing between plates are 
kept unchanged while changing the plate thickness. In order to maintain 
the approximate weight of the detector for each plate thickness, the number
of iron plates (and hence the number of RPC layers) are adjusted accordingly. 
This does not affect the analysis since the few GeV hadrons which this study 
focuses on traverse only a few layers and rarely reach the detector edges.

Pions, which constitute the major fraction of hadrons produced in neutrino
interactions in the detector mainly propagate as showers. The hits in
the shower are denoted as $x$-hits or $y$-hits depending on whether
the information originated from the $x$- or the $y$-pickup strip. The
maximum of these two numbers in a layer, named as orig-hits, is chosen
for the study. As shown in our previous work \cite{ref:moon}, the analysis
is not sensitive to this choice.

The study has been performed using fixed energy single positive pions
($\pi^{+}$). For obtaining the hit information, fixed energy single pions
are propagated from random vertices inside a volume of 200 cm $\times$
200 cm $\times$ 200 cm in the centre of the detector. This
ensures that the event is completely contained inside the detector. The
direction of propagation is determined by their zenith angle $\theta$ and
the azimuthal angle $\phi$ in this geometry which is oriented such that
the $x$-axis (the coordinate system is defined earlier) corresponds to
$\phi=0$ and the $z$-axis corresponds to the vertically up direction, with
$\theta =0$. Unless otherwise specified, $\theta$ is smeared from 0 to
$\pi$ and $\phi$ is smeared from $0$ to $2\pi$ in order to
obtain direction-averaged energy resolutions.
The pion energy is varied from 2 GeV to 15 GeV in steps of 0.25
GeV without smearing. For each energy and plate thickness, we simulate
10000 events. Eleven plate thicknesses including the default value 5.6
cm are used. The thickness is varied from 1.5 cm in steps of 0.5 cm upto
5 cm and the other thicknesses are 5.6 cm, 6 cm and 8 cm.

\begin{figure}[htb]
\centering
\includegraphics[width=0.6\textwidth]{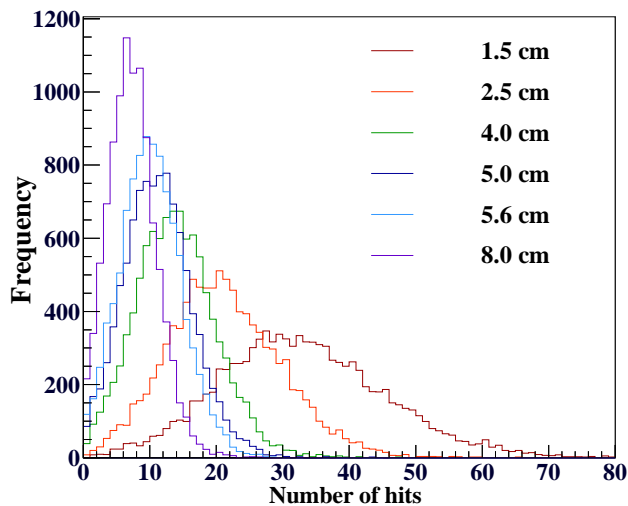}
\caption{Hit distributions of 5 GeV single pions ($\pi^{+}$) propagated
through sample iron plate thicknesses.}
\label{fig:5GeV-all}
\end{figure}

As expected, the mean number of hits increases with decreasing plate
thickness, and the width of the distribution also becomes broader.
This is illustrated in Fig.~\ref{fig:5GeV-all} which shows the hit
distributions for different iron plate thicknesses for a 5 GeV pion
($\pi^{+}$). It was found that the
magnetic field did not change the hit distribution. This is
due to the nature of shower development and multiple scattering effects
in the case of hadrons.

For the comparison of energy resolution at different thicknesses, we
choose to use the mean and rms width ($\sigma$) of the hit distributions
at different energies. We use $\sigma/E$ as the indicator of energy
resolution.

We parametrise the hadron energy resolution as $\sigma(E)/E =
\sqrt{(a^2/E+b^2)}$, where $\sigma(E)$ is the width of the distribution,
$a$ is the stochastic coefficient which depends on the thickness of
the absorber and $b$ is a constant. The analysis is done by taking
the square of the equation since it gives a linear relation between
$\left(\sigma/E\right)^{2}$ and $1/E$ with $a^2$ as the slope and $b^2$
as the intercept:

\begin{eqnarray}
\left(\frac{\sigma}{E}\right)^{2} = \frac{a^{2}}{E}+b^{2}~.
\label{eqn:sbE2a2Eb2}
\end{eqnarray}

Note that the parameters $a$ and $b$ are, in general, thickness dependent.

\section{Energy resolution for different plate thicknesses}\label{Eres}

\begin{figure}[htb]
\centering
\includegraphics[width=0.48\textwidth]{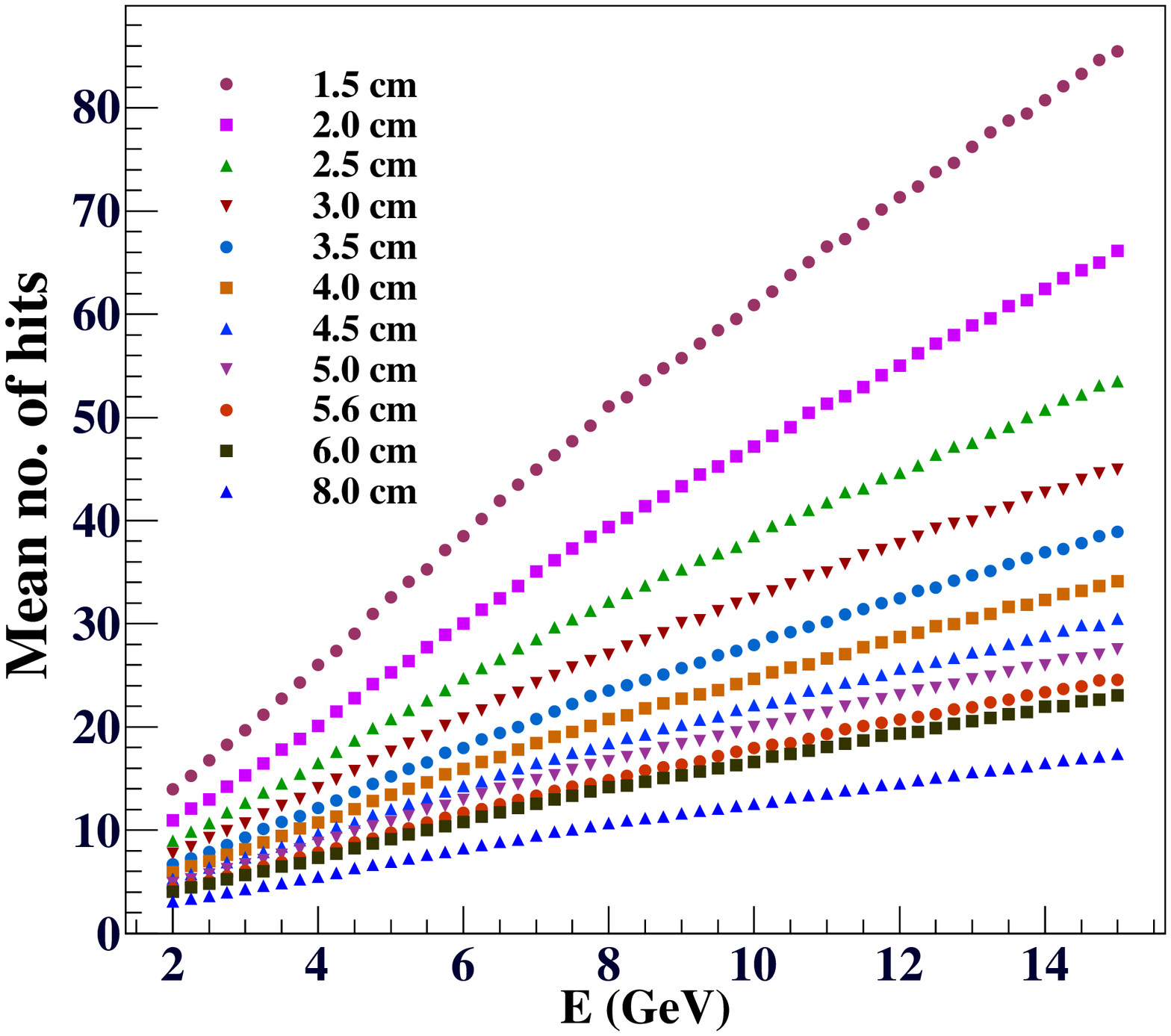}
\includegraphics[width=0.48\textwidth]{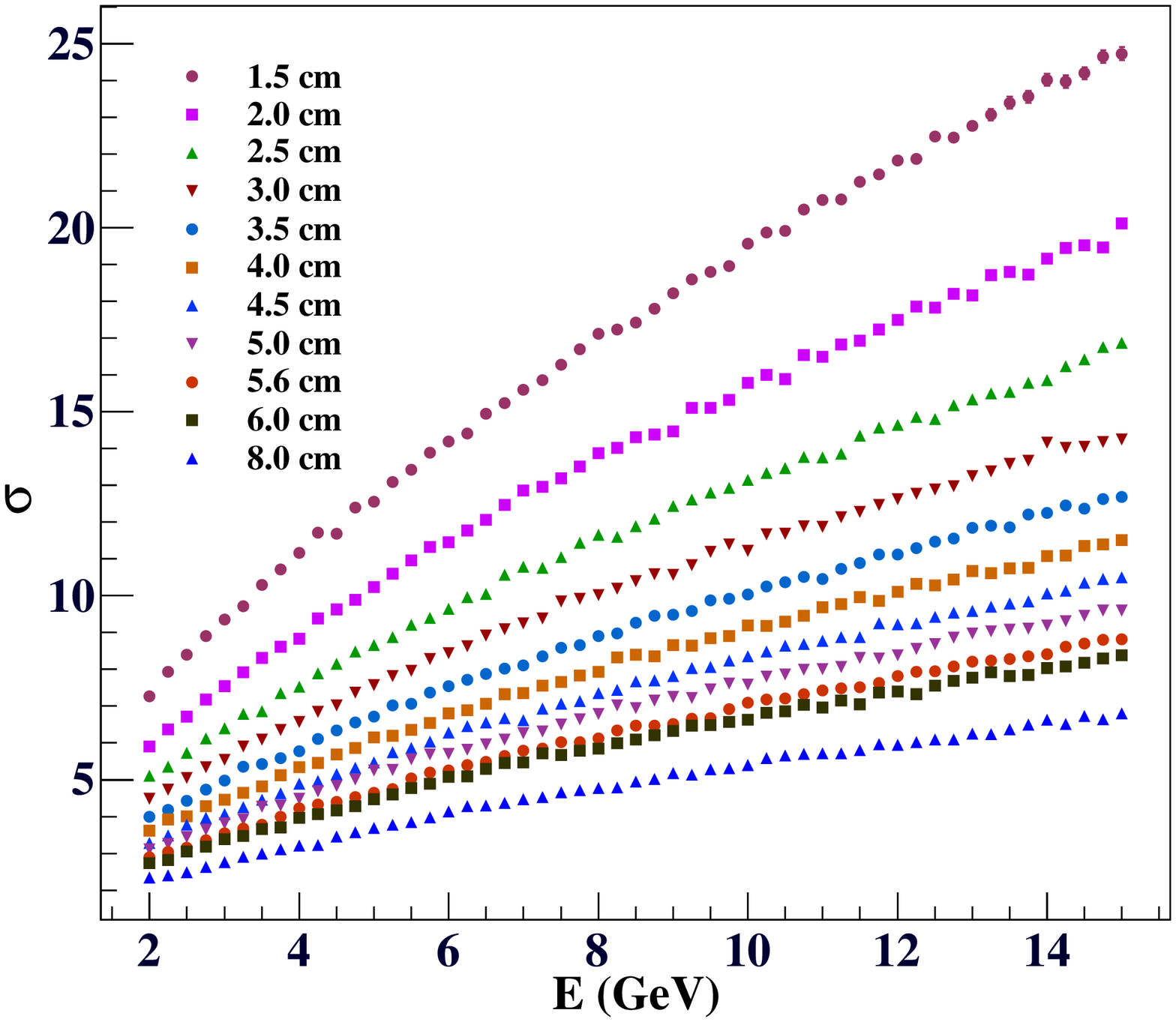}
\caption{(Left) Mean number of hits and (Right) width of hit histograms as 
functions of pion energy $E$ (GeV), in the energy range 2--15 GeV.}
\label{fig:meanhisto}
\end{figure}

Energies of interest between 2--15 GeV are used in the analyses presented
here. The arithmetic mean and rms width of the hit distributions 
in the energy range 2--15 GeV, for various thicknesses, are shown in 
Fig.~\ref{fig:meanhisto}.

In the energy region below about 5 GeV, all processes including
quasi-elastic (nucleon recoil), resonance and deep inelastic scattering
can contribute to a comparable extent to the production of hadrons in
neutrino interactions in the detector. In contrast, the high energy region
is dominated by hadrons created via deep inelastic scattering. Keeping
this distinction in mind, we analyse the response to fixed energy single
pions in two energy ranges, 2--4.75 GeV and 5--15 GeV, separately.

The ratio $\sigma/\text{Mean} = f(E, t)$ is identified as the
resolution \cite{ref:moon} and its square is fitted to the form given in
Eq.~(\ref{eqn:sbE2a2Eb2}), where $t$ is the thickness of the absorber
(iron plate) in cm (alternatively, it can be parametrised as $t/t_0$
where $t_0$ is a test or standard thickness; here $t_0 = 1$ cm).

One has to determine the specific functional form of the thickness
dependence of the parameters $a$ and $b$ on the right hand side of
Eq.~(\ref{eqn:sbE2a2Eb2}). Before doing this, it is important to determine
the values of $a$ and $b$ for different thicknesses by fitting this form
in the two different energy ranges as specified earlier.

\subsection{Energy range 2--4.75 GeV}
\begin{figure}[bth]
\centering 
\includegraphics[width=0.49\textwidth,height=0.55\textwidth]{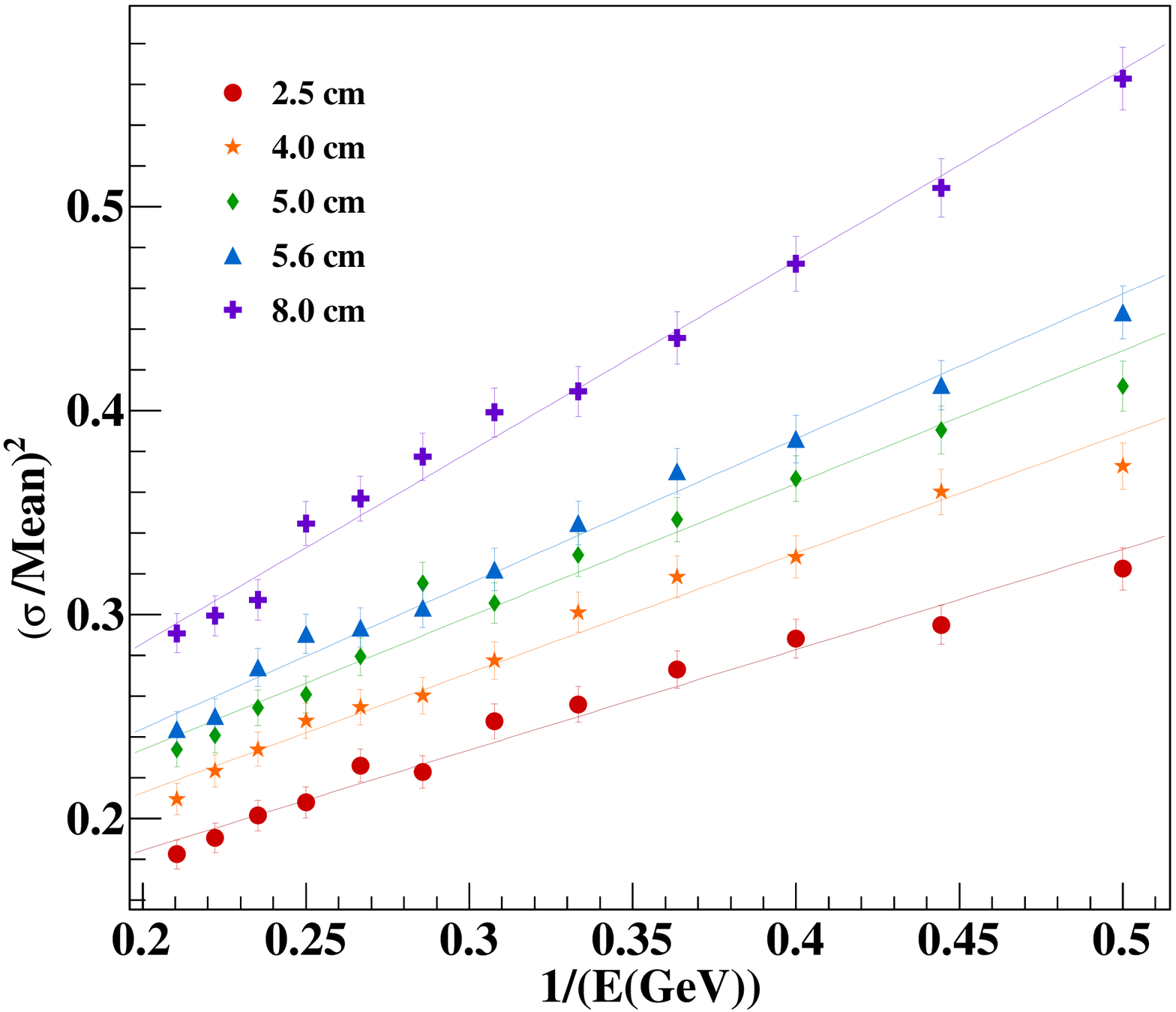} 
\includegraphics[width=0.49\textwidth,height=0.55\textwidth]{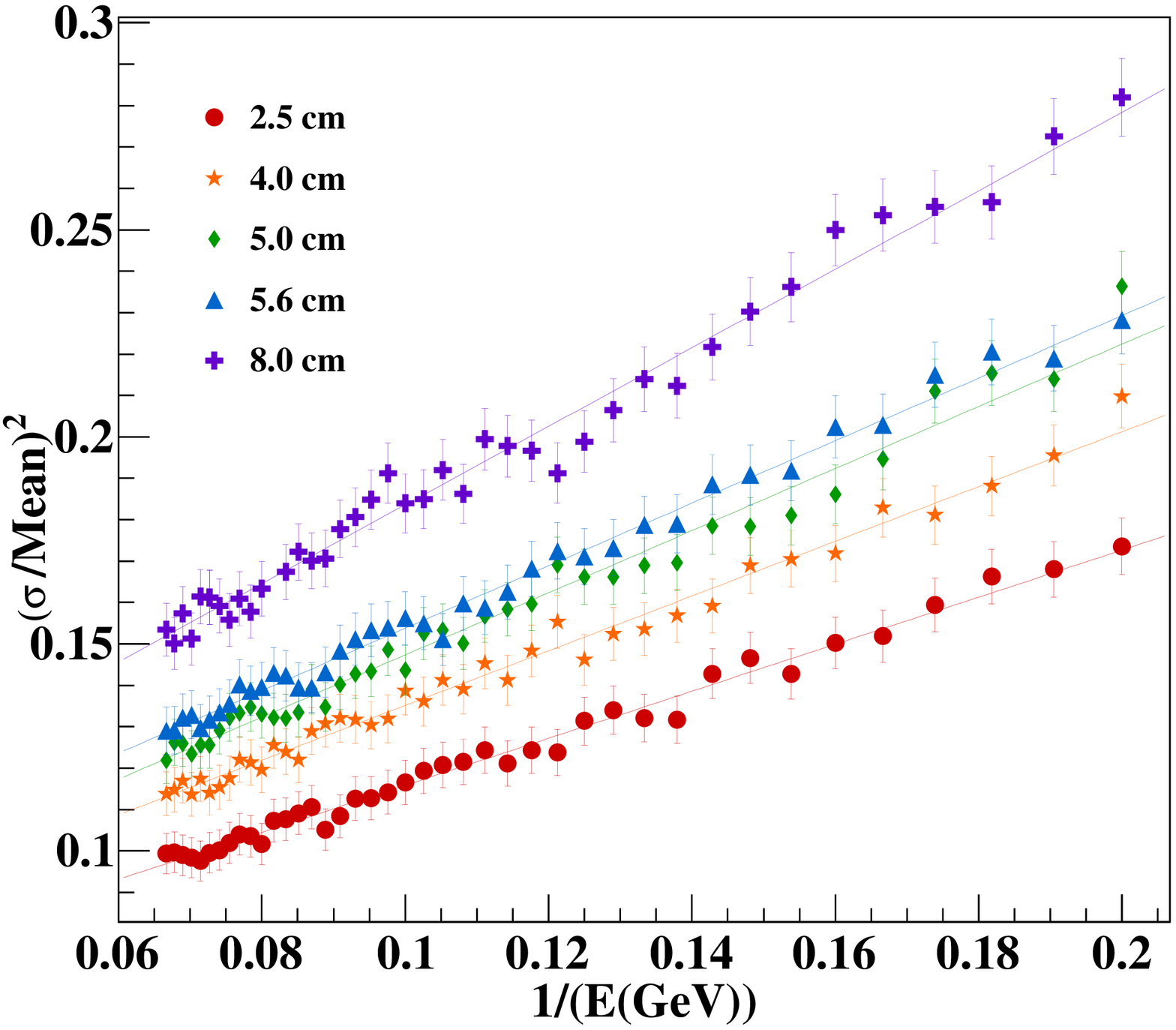}
\caption{Plots of 
$\left(\sigma/\text{Mean}\right)^{2}$ versus 
$\left(1/E(\hbox{GeV})\right)$ fitted to Eq.~(\ref{eqn:sbE2a2Eb2}) in the 
energy range 2--4.75 GeV (left panel) and 5--15 GeV (right panel). The
thickness varies from 2.5 cm to 8 cm from bottom to top.}
\label{fig:rbm2-15}
\end{figure}

We first analyze the low energy region which is the most relevant for
atmospheric neutrino studies with the
ICAL detector. The square of the resolution,
$\left(\sigma/\text{Mean}\right)^{2}$, for the thicknesses from 1.5--8.0
cm, plotted as a function of $1/E$, where $E$ is the pion energy in GeV,
is shown in Fig.~\ref{fig:rbm2-15}. It can be seen that $a$ increases
significantly with thickness as evinced by the increase in slope ($=a^2$)
of the fit with thickness, with $a$ increasing from $a=0.65$ to $a=0.97$
as the thickness increases. However, $b$, as determined by the intercept
($=b^2$), is nearly constant in comparison, as it ranges from $b=0.28$
to $b=0.31$ with increase in thickness.
\label{subsection31}

\subsection{Energy range 5--15 GeV}
\label{subsection32}

We study the higher energy region separately to probe a possible stronger
$E$-dependence. Fig.~\ref{fig:rbm2-15} shows that the behaviour is 
similar to the low energy case. Typically, the value of the stochastic 
coefficient $a$ varies from 0.70--0.97, which is higher than in the lower 
energy case by up to 10\%, whereas $b$ varies from about 0.23--0.30. 

Having determined the stochastic and constant parameters in these
different energy ranges for different thicknesses, we now proceed to
study the thickness dependence of the hadron energy resolutions.

\section{Parametrisation of the plate thickness dependence}

The functional form of the thickness dependence is introduced in
two different ways. In the first approach, the thickness dependence
is attributed entirely to the stochastic coefficient, $a$. This is
motivated by the observation that the parameter $b$ has a much smaller
dependence on the thickness, as can be seen from the analyses above. The
thickness dependence of the stochastic coefficient $a$ is parametrised
in the standard form, \begin{equation} a(t) = p_{0}t^{p_{1}}+p_{2}~,
\label{eqn:tdepreal} \end{equation} where $p_2$ is the limiting resolution
for hadrons for finite energy in the limit of very small thickness due to
the nature of their interactions, detector geometry and other systematic
effects. We estimate these parameters in suitably chosen energy ranges
as mentioned before.

We use Eq.~(\ref{eqn:tdepreal}) to determine the thickness dependence of
the stochastic coefficient $a$ separately in three different energy
ranges as mentioned earlier. The parameters $p_i~(i=0,1,2)$ are
determined independently in each energy range.

\begin{figure}[thp]
\centering
\includegraphics[width=0.6\textwidth]{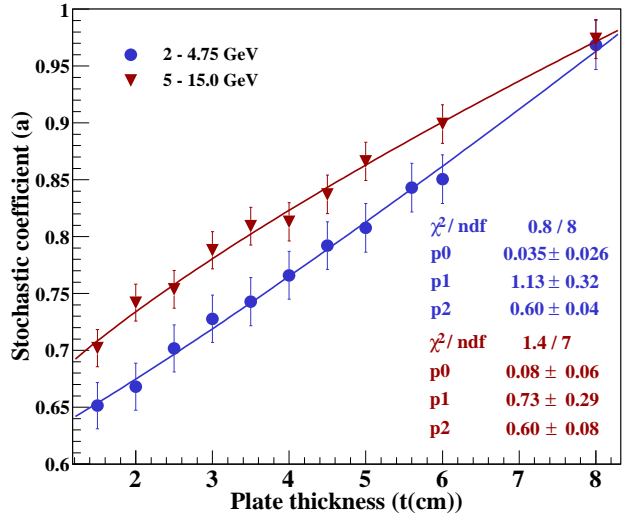}
\caption{The stochastic coefficient $a$ obtained from the analysis in
the energy ranges 2--4.75 GeV and 5--15 GeV versus
the plate thickness $t$ in cm, fitted with Eq.~(\ref{eqn:tdepreal}).}
\label{fig:tdep2-5-15}
\end{figure}

In Fig.~\ref{fig:tdep2-5-15}, we show the fits in the energy ranges
2--4.75 GeV and 5--15 GeV as functions of thickness. The parameters
$p_0$, $p_1$, and $p_2$ obtained from the fit to the form given in
Eq.~(\ref{eqn:tdepreal}) are also shown in the figure. The thickness
dependence is given by the exponent $p_1$. From the fit value shown in 
Fig.~\ref{fig:tdep2-5-15}, $p_1$ is clearly energy-sensitive and 
decreases in the higher energy range.

The analyses followed in the two energy ranges show that the dependence 
on the thickness is stronger than $\sqrt{t}$; however, note the smallness
of $p_0$, the coefficient of the thickness parameter, in all cases,
in comparison with the constant parameter $p_2$. Irrespective of the
energy range, it remains around $p_2 \sim 0.60 \pm 0.08$ and contributes
substantially to the resolution. This means that {\it there will always
be a residual resolution which cannot be improved further by reducing
the thickness}, thereby making the option of going to smaller thicknesses
less attractive than what the bare $t$-dependence indicates. For example,
although the resolution has an approximately linear dependence on the
thickness ($p_1 \sim 1.1$) at low energy, it worsens by only about 15\%
when the thickness doubles from $t=2.5$ to $t=5$ cm rather than doubling
as the bare $t$ dependence indicates.

\begin{figure}[thp]
\centering
\includegraphics[width=0.6\textwidth]{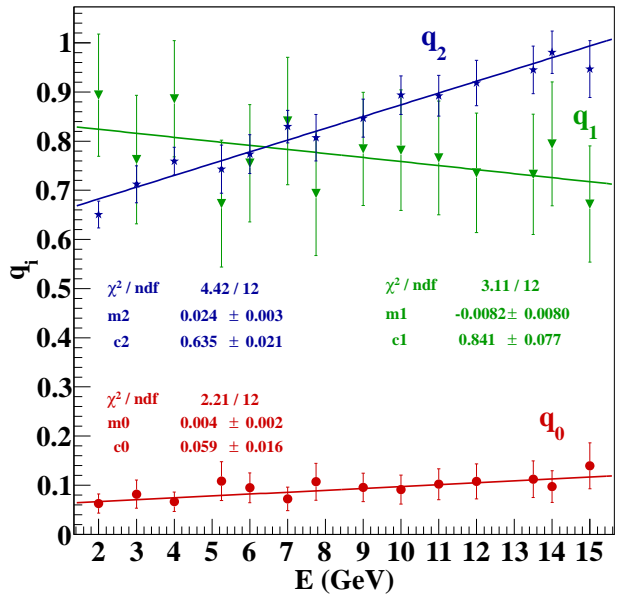}
\caption{Variation of the parameters $q_{i}$ obtained by fitting
$\sigma/\sqrt{E}$ to Eq.~(\ref{eqn:tdepreal2}) in the energy range 2--15
GeV. The linear fits through the points indicate the $E$ dependence for
each parameter.
}
\label{pi-trends}
\end{figure}

An alternative approach is to analyse the thickness dependence of
the entire width and not just that of the stochastic coefficient. The
analysis was done for different energies. A fit to $\sigma/\sqrt{E}$
with the equation
\begin{equation}
\sigma/\sqrt{E} = q_{0}t^{q_{1}}+q_{2}~, 
\label{eqn:tdepreal2}
\end{equation} 
congruent in form with Eq.~(\ref{eqn:tdepreal}), reveals the following
trend as illustrated in Fig.~\ref{pi-trends}. The exponent $q_{1}$ of the
absorber thickness (t(cm)) decreases from $\sim$ 0.9 to 0.66 in the 2--15
GeV energy range, whereas its coefficient $q_{0}$ increases from $\sim$
0.06--0.14 with energy. The constant term $q_{2}$ increases from $\sim$
0.65--0.98 with energy $E$ (GeV). Again, the smallness of the coefficient
$q_0$ results in the $q_2$ dominating over the term $q_0t^{q_1}$. Thus
the behaviour closely parallels that of the earlier analysis with
the thickness dependence of $a$ alone. We also show linear fits for the
$E$ dependence of the parameters $q_0, q_1$ and $q_2$ in
Fig~\ref{pi-trends}. The
trends indicate that the thickness exponent mildly decreases with energy
and may therefore be compatible with the square-root results of earlier
studies at higher beam energies \cite{ref:green,ref:numi0335}.

We have focussed here on single pions since we are
interested in understanding the thickness dependence. However, ICAL will
be built to study charged-current neutrino interactions where multiple
hadrons may be produced in the final state, from either resonance or
deep inelastic interactions. A brief remark about hadrons from neutrino
interactions is therefore in order. The study of such final states
where the hadronic energy is shared by more than one hadron will bring
additional uncertainty in the study of thickness dependence.

An earlier study \cite{ref:moon} for a fixed absorber thickness of $t=5.6$ cm
compared single pion resolutions with those of events generated by
the NUANCE neutrino generator \cite{nuance} where multiple hadrons are
produced in the final state and there is a non-trivial partition of energy
into the different hadronic final states. The trends in the dependence
of the resolution (quantified by $\sigma/E$ as well as the stochastic
coefficient $a$) were found to be similar in the single-pion and
multi-hadron cases. This can be understood from the $e/h$ response of
ICAL; details are given in Appendix A. Again, we find that the thickness
dependence from the NUANCE data sample has a similar behaviour to the
single pion case.

The effect of different hadron models on resolutions were
also studied by replacing the LHEP model that was used in GEANT4
\cite{ref:geant4} with QGSP (for 12 GeV hadrons) and QGSP\_BERT (for
the lower energy hadrons). The resolution was found to be reasonably
model independent, with a variation in the mean (rms) of less than 4\%
(5\%) among different models in the energy range from 2--15
GeV for $t=5.6$ cm.

\section{Angular dependence of energy resolution and thickness
dependence}
 
\begin{figure}[hbp]
\centering
\includegraphics[width=0.55\textwidth]{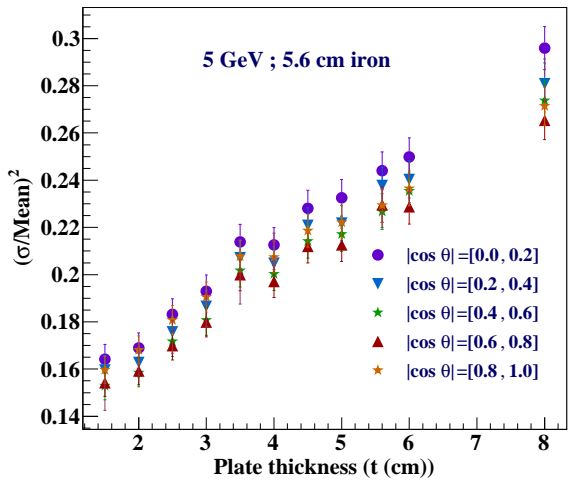}
\caption{Energy response of a 5 GeV pion as a function of various
thicknesses in different incident $\cos\theta$ bins.}
\label{5GeV}
\end{figure}

So far, we have considered hadrons smeared in all
directions in both polar and azimuthal angles, $\theta$ and $\phi$. In
this section we present the energy resolution in various bins of
incident polar angle $\theta$. The azimuthal angle $\phi$ is still
smeared from 0--2$\pi$ in each $\theta$ bin. The bins are defined
symmetrically over the up/down directions in intervals of 0.2 in
$\vert\cos\theta\vert$, with the averages corresponding
to $\langle \vert\cos\theta_{in}\vert\rangle $ = 0.9, 0.7, 0.5, 0.3
and 0.1 respectively. The bins with the largest $\vert \cos\theta \vert$
are nearly perpendicular to the iron plates (and we refer to them as
vertical events) while the ones with the smallest values are practically
parallel to them (and we label them as horizontal events). As an example,
the energy resolution for a 5 GeV pion in different angular bins as a
function of plate thickness ($t$ (cm)) is shown in Fig.~\ref{5GeV}.

Several observations are in order. Firstly, hadrons
in the horizontal bin have the worst resolution, as expected. The
resolution generically improves with increasing $\vert \cos\theta
\vert$, except in the vertical bin ($\vert\cos\theta\vert = 0.8$--1.0)
for all thicknesses. This is due to the geometry of the detector, with
support structures at every 2~m in both the $x$- and $y$-directions. This
reduces the region of sensitive detector in the vertical direction, with
a consequent loss of resolution. Finally, while the hadron traverses an
effective thickness of $t/\cos\theta$, the resolution does not exhibit
such a naive scaling behaviour. It is seen that, for the same value of
$t/\cos\theta$, the resolution is better at smaller thicknesses than at
larger thicknesses. This is again because of the non-trivial geometry
and other factors.

\begin{figure}[thp]
\centering
\includegraphics[width=0.45\textwidth,height=0.5\textwidth]{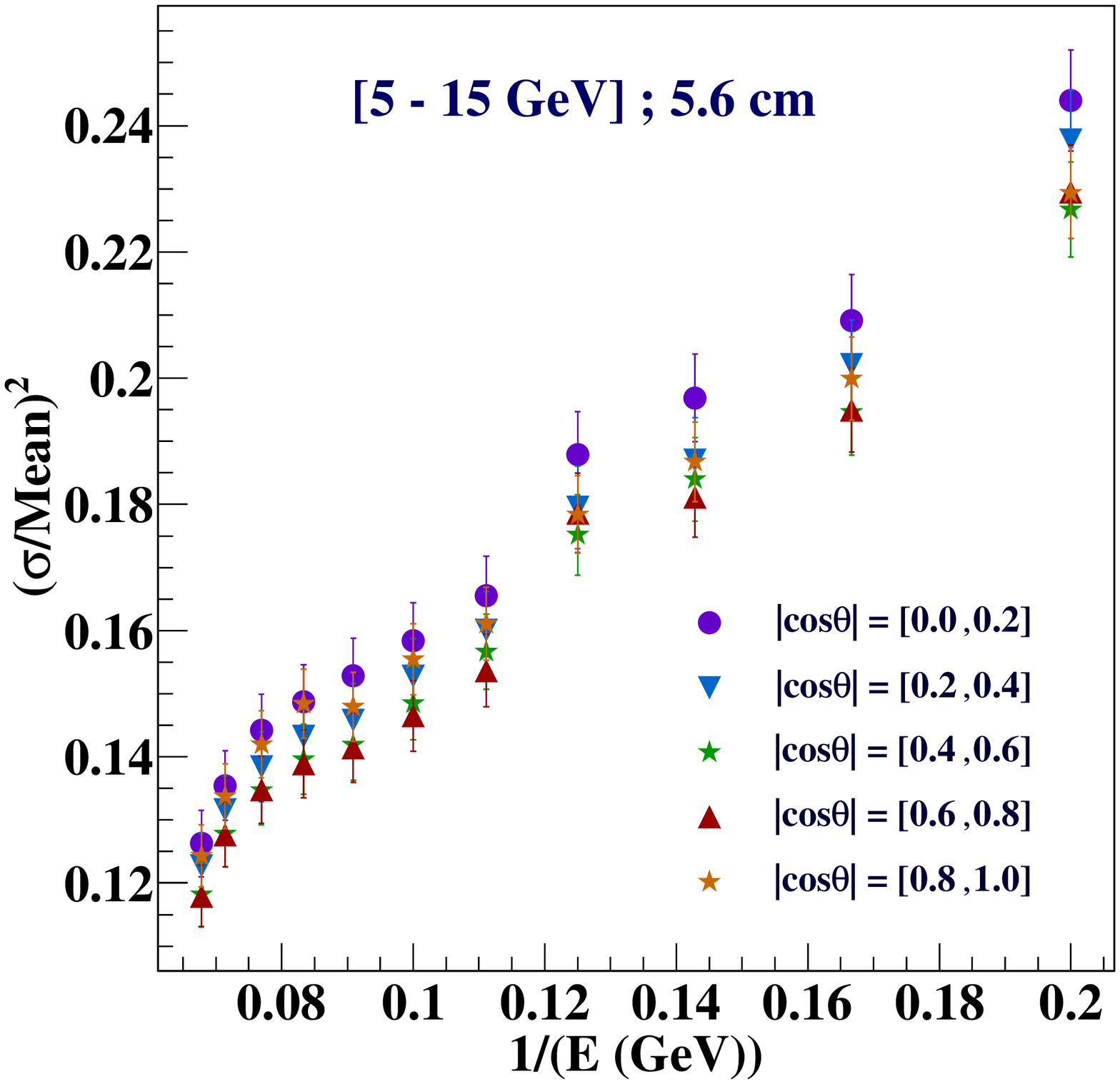}
\includegraphics[width=0.45\textwidth,height=0.5\textwidth]{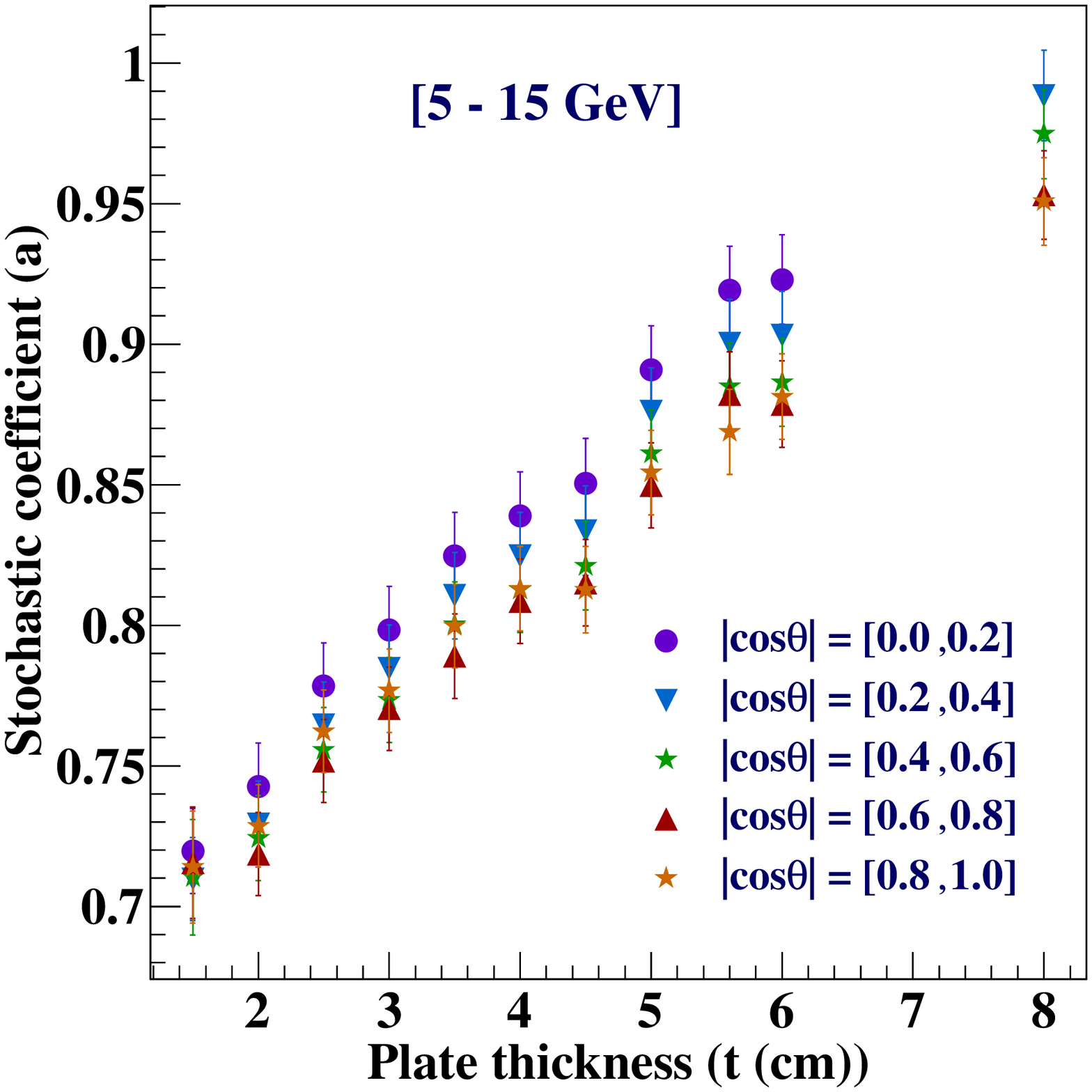}
\caption{(Left) Energy resolution in bins of incident $\theta$ in the energy
range 5--15 GeV for the default iron thickness of $t=5.6$ cm.
(Right) Stochastic term $a$ vs plate thickness in various incident $\theta$
bins for the energy range 5--15 GeV.}
\label{5.6cm-A-bin}
\end{figure}

A similar trend is seen at all energies; the energy resolution
in each $\cos\theta$ bin, in the energy range 5--15 GeV, is shown in
Fig.~\ref{5.6cm-A-bin} for the default thickness of $t=5.6$ cm. The
thickness dependence in the angular bins is shown most conveniently in
terms of the stochastic coefficient $a$ as determined from fits in the
energy range 5--15 GeV and has also been plotted in
Fig.~\ref{5.6cm-A-bin}. 

Since there is only a mild dependence on the hadron direction, the
direction-averaged results obtained in the earlier sections present a
realistic picture of the thickness dependence of the energy resolution
of hadrons.

\section{Comparison with other experiments}

For the validation of the analysis presented here, the result of ICAL simulation 
has been compared with both simulations of MONOLITH and MINOS collaborations
and also to the data from the test beam runs conducted by them. In
the current simulation, ICAL with 8 cm iron plate has a resolution
of 98.5\%$/\sqrt{E}\oplus 29.4$\% which is roughly comparable to the
angle-averaged result of 90\%$/\sqrt{E}\oplus 30$\% obtained from the
simulation studies of MONOLITH \cite{ref:monolith} with the same plate
thickness. For convenience of comparison, the convention $a/\sqrt{E}
\oplus b \equiv \sqrt{a^2/E + b^2}$ has been used.

Our results cannot be directly compared with the test beam data
since beams are highly directional (with $\cos\theta = 1$ as the beam
divergence is typically small). To enable a comparison with the
beam data we consider events where the hadrons are normally incident
on the detector plates.

\begin{figure}[htb]
\centering
\includegraphics[width=0.55\textwidth]{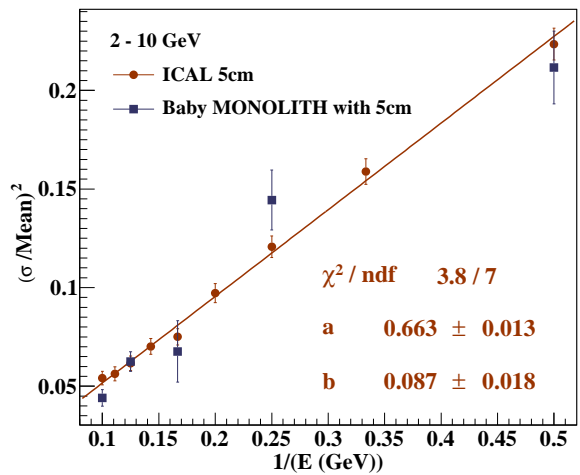}
\caption{Energy response of ICAL detector with 5 cm thick iron plates
with single pions in the energy range 2--10 GeV, propagated from the
vertex fixed at $(100, 100, 0)$ cm in the vertical direction, 
compared with the data from MONOLITH test beam run
\protect\cite{ref:babymonolith,ref:t7-ps-beam-run}.}
\label{ical-5cm-fv}
\end{figure}

MONOLITH has performed a test beam run with 5 cm
iron plates (Baby MONOLITH) with the T7-PS beam at CERN
\cite{ref:babymonolith,ref:t7-ps-beam-run}. This beam provides pions of
energies ranging from 2--10 GeV which are exactly normally incident on the
iron plates. The run reported an energy resolution of $68\%/\sqrt{E}\pm
2\%$. The simulation of ICAL detector with 5 cm iron plates with single
pions of energies 2--10 GeV incident normally on the detector at a fixed
vertex $(100, 100, 0)$ cm, is shown in Fig.~\ref{ical-5cm-fv}. The analysis
gives a similar energy resolution of $66.3\%/\sqrt{E}\oplus 8.7\%$.

The simulation studies with gaseous detectors by MINOS collaboration have
reported a hadron energy resolution of 70\%/$\sqrt{E}$ with 1.5$''$ (i.e.,
3.8 cm) iron plates \cite{ref:petyt}. The test beam run of MINOS with APT
(Aluminium Proportional Tubes) active detectors and 1.5$''$ steel plates
in the energy range 2.5--30 GeV was reported to have a hadron energy
resolution of 71\%/$\sqrt{E} \oplus 6$\% \cite{ref:numi0335}. ICAL
simulation with 4 cm iron plates in the energy range 2--30 GeV
gives a resolution of 61\%/$\sqrt{E} \oplus 14$\%. The results are
compatible, considering that there are differences in detector
geometries.

Our simulation results thus agree with those of MINOS and MONOLITH simulations and test
beam results within statistical errors. The slight
differences can be attributed to differences in the details of the
detector configuration. Note also that fixed vertex data tend to give
smaller values of $b$ than the smeared vertex case; this is because the
hadrons see more inhomogeneities in the detector geometry in the latter
case, and this is reflected in the larger residual resolution.

\section{Conclusions}

We have made a simulation study of the direction-averaged hadron energy resolution as a
function of iron plate thickness (from 1.5 to 8 cm) in the energy range of
interest for atmospheric neutrino interactions. The study was motivated
by the realisation that the hadron energy resolution is a crucial
limiting factor in reconstructing the neutrino energy in charged current
interactions of atmospheric neutrinos in the magnetized Iron CALorimeter
(ICAL) detector at the proposed India-based Neutrino Observatory (INO).
The analysis was done by propagating pions in the simulated ICAL detector
at various fixed energies, averaged over all directions $(\theta,\phi)$
in each case.

Simulations show that the hadron energy resolution depends on plate
thickness $t$ (cm) through a relation $a = p_0\, t^{p_1}+p_2$, where $a$,
the stochastic coefficient, is the energy-dependent term in the standard
resolution, $(\sigma/E)^2 = a^2/E + b^2$. That is, there is a finite 
energy resolution for hadrons even when the plate thickness is small. This
reflects the strong nature of hadronic interactions with matter (iron
in this case) that leads to large systematic uncertainties. We find that 
the constant term $p_2$ is always dominant compared to the first
$t$-dependent term because the coefficient $p_0$ of the $t^{p_1}$
term is small; hence reducing the plate thickness does not lead to a
significant gain in the hadron energy resolution. This is true over
all the thicknesses studied in the energy range 2--15 GeV.

Similar results are obtained when the quantity $\sigma/\sqrt{E} = q_0
t^{q_1} + q_2$ is studied for its thickness dependence. The trends of
the fit parameters $q_{i}$ as functions of $E$ (GeV) show that the
the smallness of the coefficient $q_0$ is again responsible for the
dominance of the thickness independent term $q_{2}$. The results were
also reasonably insensitive to the choice of hadron model within GEANT.

The energy resolution was also studied as a function of the incident
angle $\theta$ for different thicknesses $t$. It was found that
the thickness dependence of the resolution did not satisfy the naive
expectation of being proportional to $t/\cos\theta$; this is because of
the detector geometry with the distribution of its sensitive elements
also playing a role. On the whole, the angular dependence is not strong.
While the resolution did improve with increasing $\vert\cos\theta\vert$
as expected, the vertical hadrons were not as well-resolved as would have
been the naive expectation because of the presence of dead spaces such
as detector support structures, etc.

Comparisons of ICAL simulations with those of MONOLITH and MINOS and
their test beam runs have been conducted and are found to match.

The final choice of the plate thickness will depend not only on the behaviour
of hadrons but also on on the energy range of interest to the physics
goals of the experiment. Issues like low energy muons, the threshold
energy, possibility of electron detection, cost etc will also affect the
choice of plate thickness. But these are outside the scope of this simulation
study.

\section*{Acknowledgement}

We thank G.~Majumder, A.~Redij for the development of the simulations
framework for ICAL; B.~Choudhary, V.M.~Datar and Y.P.~Viyogi for the
critical comments and suggestions on the manuscript and content. We also
thank N.K. Mondal, P.~K Behera and all members of the INO collaboration
who gave their comments and suggestions in the weekly simulations
meetings. We thank the Department of Atomic Energy (DAE), India, and
Department of Science and Technology (DST), India, for supporting this
research work.

% \appendix
\section*{Appendix A: Study of $e/h$ ratio in the context of plate thickness dependence}
\label{sec:appA}
A hadron shower consists of both hadronic and electromagnetic parts. The
electromagnetic part of a hadron shower originates from the neutral pions
$\pi^{0}$. The response of neutral pions is similar to that of electrons
since the $\pi^0$ decays almost immediately into electron-positron pairs.
Hence a study of the ratio of electron response to charged pion response,
i.e, the $e/h$ ratio, helps us characterise the effect of neutral hadrons
on the energy resolution.

\begin{figure}[bhp]
%   \centering
\includegraphics[width=0.49\textwidth]{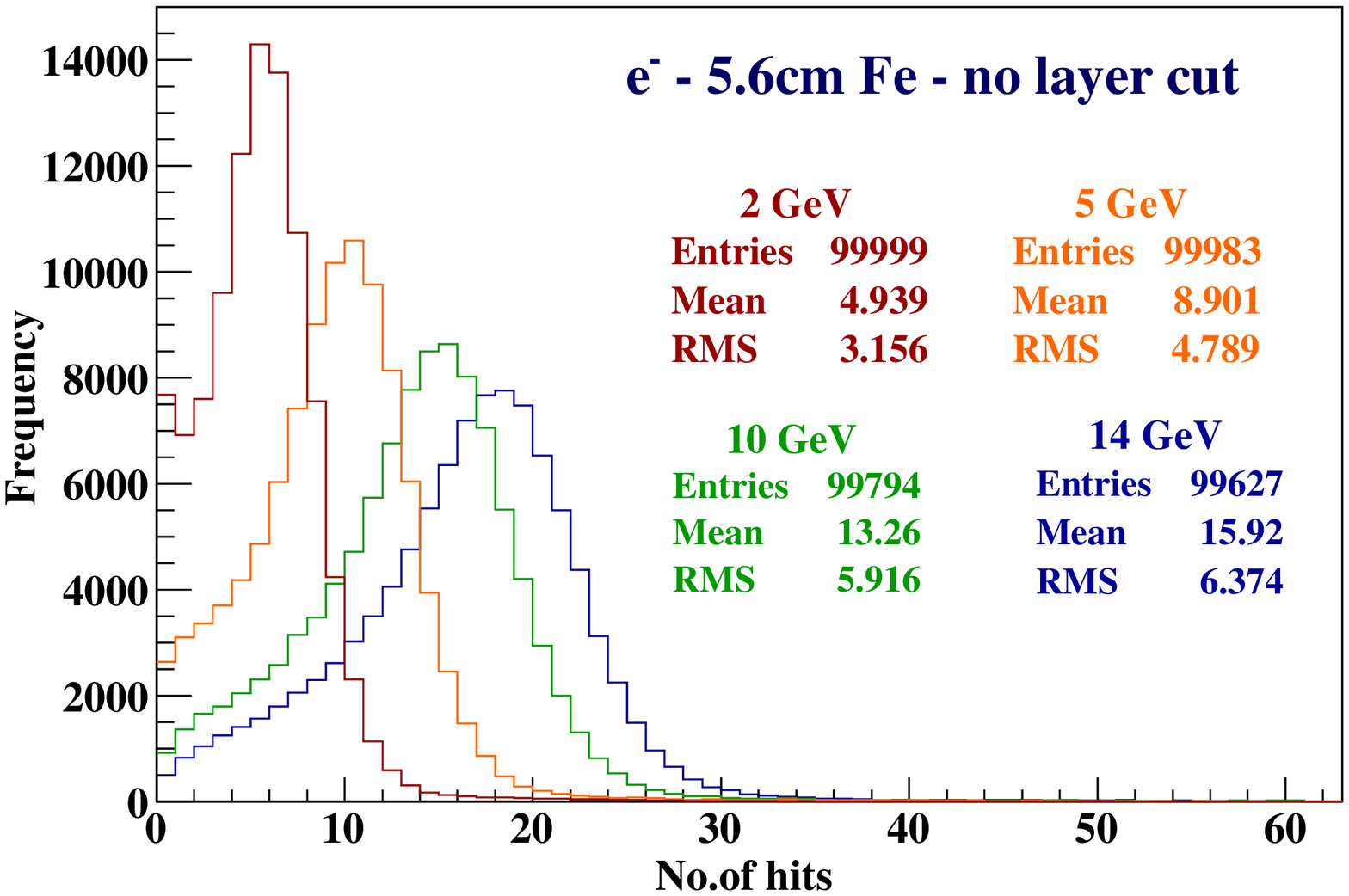}
\includegraphics[width=0.49\textwidth]{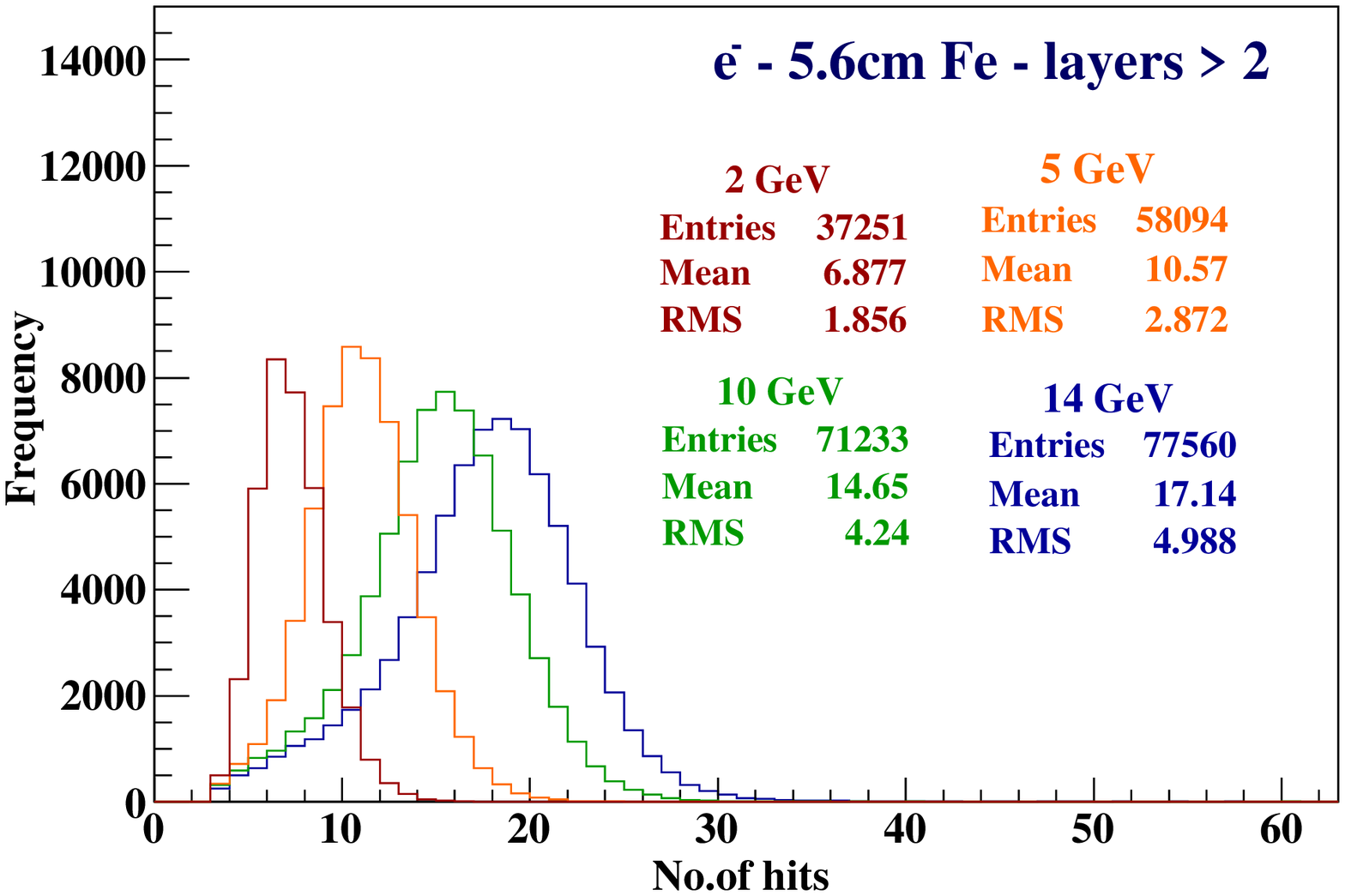}

\includegraphics[width=0.49\textwidth]{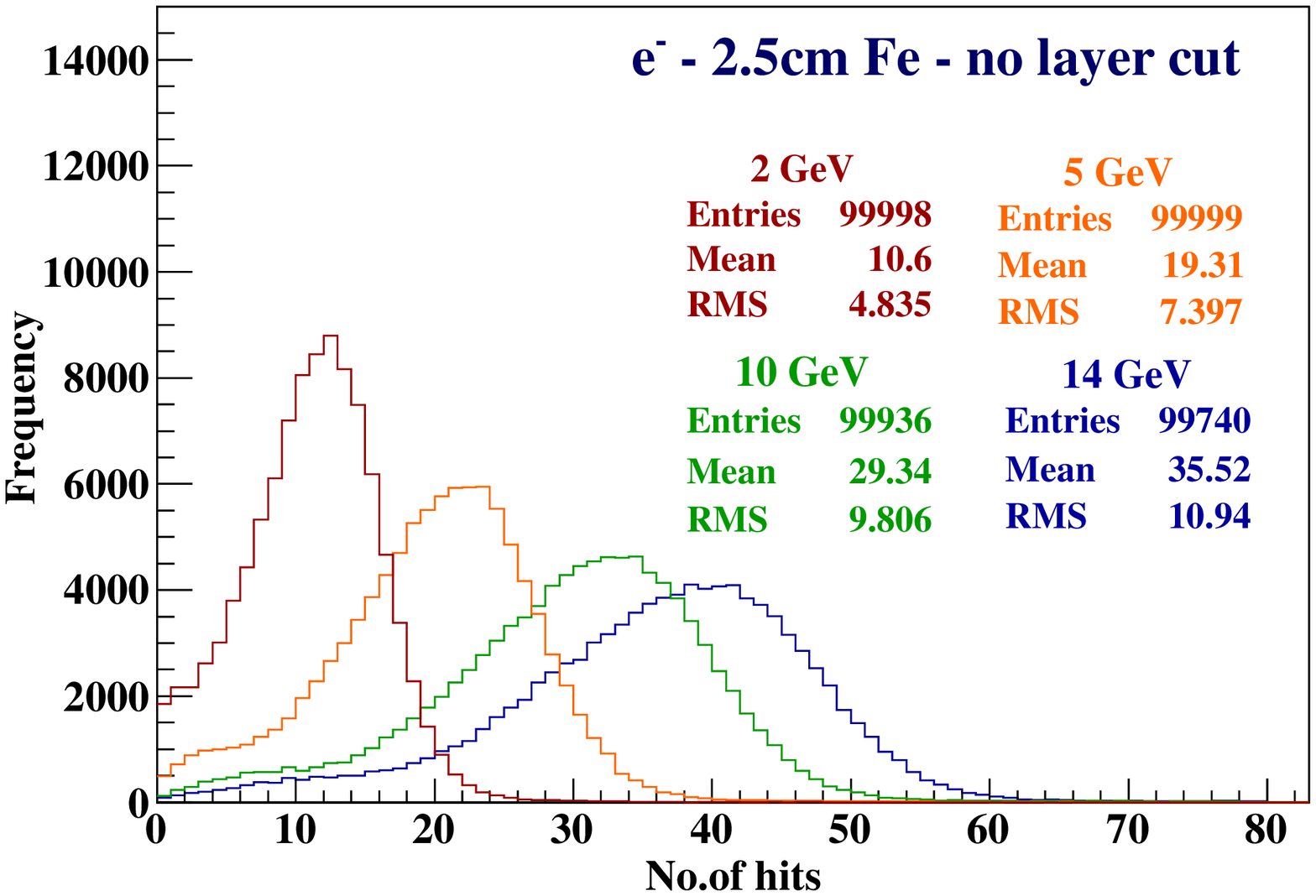}
\includegraphics[width=0.49\textwidth]{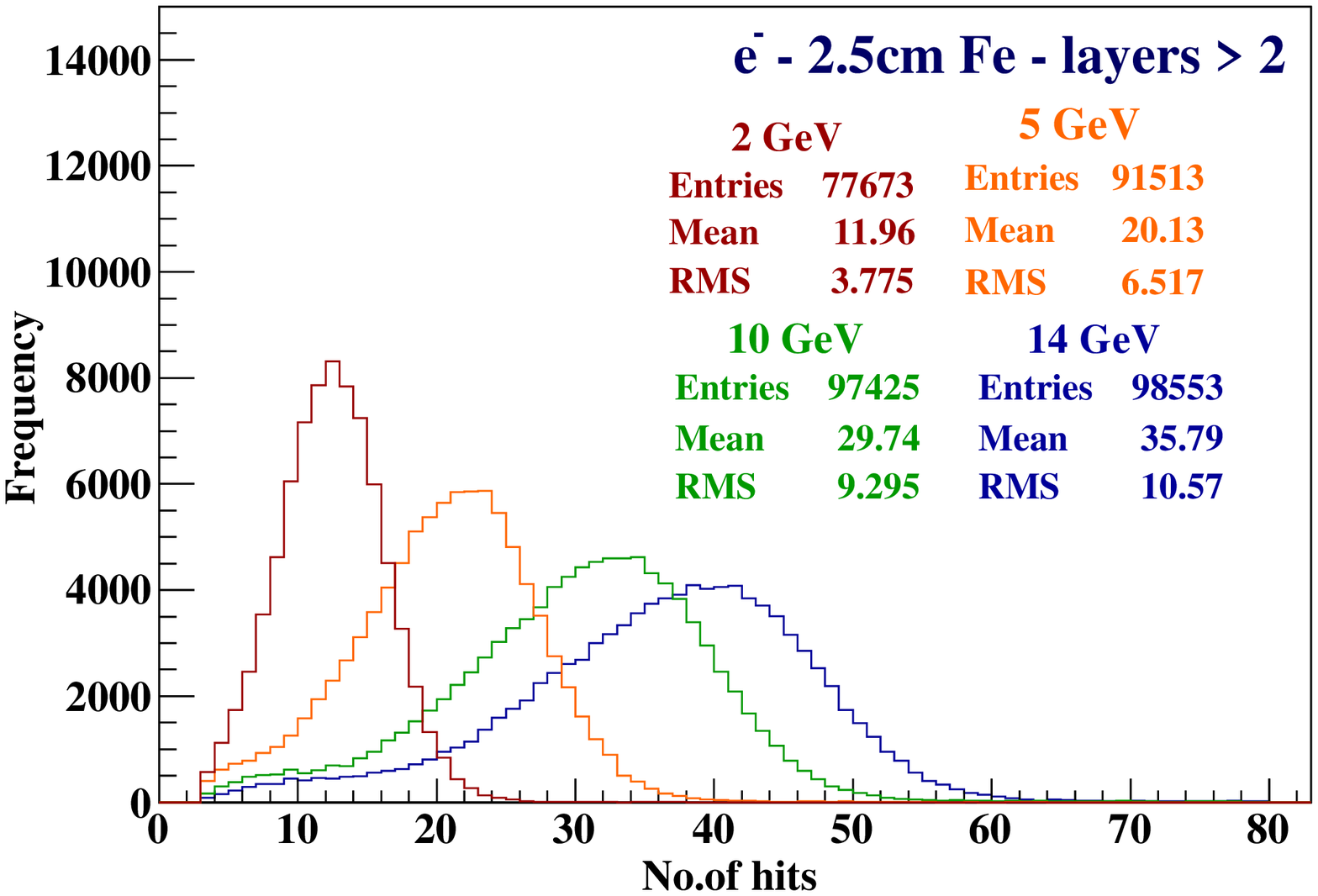}
\caption{Hit distributions of fixed energy single electrons at 2, 5,
10 and 14 GeV in 5.6 cm (top) and 2.5 cm (bottom) thick iron, averaged over
all directions. The left panels show the distributions without any layer
cut and the right panels those with the layer cut of $l > 2$, where
$l$ is the number of layers containing hits.
}
\label{e-hit-histos}
\end{figure}

We have conducted the simulation studies of the $e/h$ ratio in ICAL
detector using fixed energy single electrons and pions. Here, 1,00,000
single particles, this time electrons, are generated in the energy range
2--15 GeV and are propagated in arbitrary directions (with $\theta$
smeared from $0-\pi$ and $\phi$ from $0-2\pi$) within a volume of 2
m $\times$ 2 m $\times$ 2 m in the central region of the ICAL detector
for different iron plate thicknesses. The response is very smooth as
a function of thickness; hence the results only for 5.6 cm and 2.5 cm
are shown in Fig.~\ref{e-hit-histos}. The hit distributions averaged
over all directions for $2, 5, 10$ and 14 GeV electrons in the two
sample iron plate thicknesses 2.5 cm and 5.6 cm are illustrated in
Fig.~\ref{e-hit-histos}. For reference, the corresponding pion hit
distributions at 5 GeV have been illustrated in Fig.~\ref{fig:5GeV-all}.

It can be seen that the peak positions are not very different between
electrons and pions; however, there are many more zero hits in the former
especially at the higher thickness of 5.6 cm due to the large energy
loss of electrons in the iron. This can be improved by imposing a
selection criterion that hits must be present in at least 3 layers in
each event. As can be seen in the
right hand panels of Fig.~\ref{e-hit-histos}, the histograms are then
more symmetrical about the peak, which is now shifted to the
right. In addition, this criterion primarily affects results at higher
thicknesses and lower energies, where the layer cut improves the hit
distribution significantly although the reconstruction efficiency is
reduced. As the thickness reduces, the layer cut does not affect the
hit distribution significantly, except at low energies below about 5 GeV.

\begin{figure}[hbp]
 \centering
\includegraphics[scale=0.4]{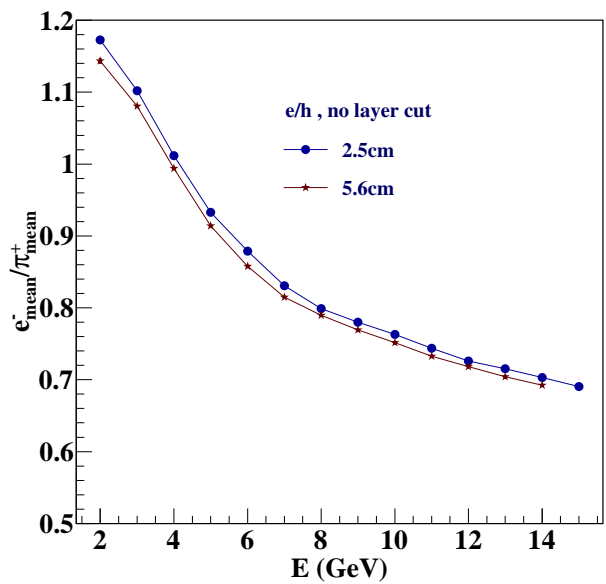}
\caption{Variation of $e/h$ ratio with incident energy for two different
thicknesses namely 2.5 cm and 5.6 cm without any layer cut. It can be
seen that the $e/h$ ratio decreases with increase in energy.}
\label{e-by-h}
\end{figure}

The electron energy is calibrated to the mean number of hits as in the
case of fixed energy single pions. The ratio of the electron response to
charged pion response, i.e., the $e/h$ ratio, is obtained as:
\begin{equation}
e/h = e^-_{mean}/{\pi^{+}_{mean}},
\end{equation}
where $e^-_{mean}$ is the arithmetic mean of the electron hit distribution
and ${\pi^{+}_{mean}}$ is the arithmetic mean of the hit distribution
for $\pi^{+}$. If $e/h$ = 1, then the detector is said to be compensating.
The variation of the $e/h$ ratio with incident energy for the two
sample thicknesses 2.5 cm and 5.6 cm are shown in Fig~\ref{e-by-h}.

It can be seen that the value of $e/h$ decreases with energy.  However,
it should be noted that there is no direct measurement of the energy
deposited in ICAL.  Here the energy of a shower is simply {\it calibrated}
to the number of hits, and electrons which travel smaller distances in a
high Z material like iron have lower number of hits compared to charged
pions. At lower energies the electron shower hits are concentrated
around a small region. At these energies, the charged pions also do
not traverse many layers in the detector due to the larger hadronic
interaction length. This causes the mean of the electron hit distribution
to be roughly the same or slightly larger than that of the $\pi^{+}$
hit distribution. But with the increase in energy, the charged pions
travel more distance and hence give more hits (as they traverse more
layers) since the hadronic interaction length is much more than the
electromagnetic interaction length at higher energies and hence the ratio
of hits in the two cases drops with energy. The layer cut only affects
the low energy result by marginally decreasing the $e/h$ ratio at $E <
4$ GeV, that too only for higher thicknesses.

In a neutrino interaction where all types of hadrons can be produced
(although the dominant hadrons in the jet are pions), the response of
ICAL to hadrons produced in the interaction depends on the relative
fractions of charged and neutral pions. The NUANCE neutrino
generator was used to generate charged current atmospheric muon neutrino
events for the default ICAL thickness of 5.6cm. The fraction of the
different types of hadrons obtained from a 100 year sample was found
to be $\pi^+: \pi^-: \pi^0::0.38:0.25:0.34$, with the remaining 3\%
contribution mainly from kaons.

The average response of hadrons obtained from the charged current muon
neutrino interaction can be expressed as:
\begin{eqnarray}
 R_{had} & = & \left[(1-F_0) \times h + F_0 \times e \right], \\
 \nonumber
         & = & h \left[(1-F_0) + F_0 \times \frac{e}{h} \right],
 \label{eqn-e-by-pi}
\end{eqnarray}
where $e$ is the electron response, $h$ the charged hadron response and
$F_0$ is the neutral pion fraction in the sample.

The atmospheric neutrino events of interest in ICAL are dominated by
low energy events with hadrons typically having energies $E < 10$
GeV for which the average value of $e/h$ is $e/h\approx0.9$. Using
$F_0~=~0.34$ in Eq.~(\ref{eqn-e-by-pi}), we get the average hadron response
for NUANCE-generated events to be $R_{had} = 0.97 h$ which is not very
different from $h$. For this reason, the analysis of response with
multiple hadrons in NUANCE-generated events sample was not very different
from that of the single pions sample, as discussed in Ref.~\cite{ref:moon}.

\end{document}